\documentclass[preprint]{elsarticle}

\usepackage{lineno,hyperref}

\usepackage{amsmath}
\usepackage{amssymb}
\usepackage{amsthm}

\usepackage{algorithm}
\usepackage{algorithmic}

\journal{Knowledge-Based Systems}

\begin{document}

\begin{frontmatter}

\title{Mining Application-aware Community Organization with Expanded Feature Subspaces from Concerned Attributes in Social Networks
}


\author[mymainaddress,mysecondaryaddress]{Peng Wu}
\ead{catking@sjtu.edu.cn}

\author[mymainaddress,mysecondaryaddress]{Li Pan\corref{mycorrespondingauthor}}
\cortext[mycorrespondingauthor]{Corresponding author}
\ead{panli@sjtu.edu.cn}

\address[mymainaddress]{School of Electronic Information and Electrical Engineering, Shanghai Jiao Tong University, 800 Dong Chuan Rd, Shanghai, China}
\address[mysecondaryaddress]{National Engineering Laboratory for Information Content Analysis Technology, Shanghai Jiao Tong University, Shanghai, China}

\begin{abstract}
Social networks are typical attributed networks with node attributes. Different from traditional attribute community detection problem aiming at obtaining the whole set of communities in the network, we study an application-oriented problem of mining an application-aware community organization with respect to specific concerned attributes. The concerned attributes are designated based on the requirements of any application by a user in advance. The application-aware community organization w.r.t. concerned attributes consists of the communities with feature subspaces containing these concerned attributes. Besides concerned attributes, feature subspace of each required community may contain some other relevant attributes. All relevant attributes of a feature subspace jointly describe and determine the community embedded in such subspace. Thus the problem includes two subproblems, i.e., how to expand the set of concerned attributes to complete feature subspaces and how to mine the communities embedded in the expanded subspaces. Two subproblems are jointly solved by optimizing a quality function called subspace fitness. An algorithm called ACM is proposed. In order to locate the communities potentially belonging to the application-aware community organization, cohesive parts of a network backbone composed of nodes with similar concerned attributes are detected and set as the community seeds. The set of concerned attributes is set as the initial subspace for all community seeds. Then each community seed and its attribute subspace are adjusted iteratively to optimize the subspace fitness. Extensive experiments on synthetic datasets demonstrate the effectiveness and efficiency of our method and applications on real-world networks show its application values.
\end{abstract}


\end{frontmatter}


\section{Introduction}
Community structure is one of the most prominent features of social networks, as it helps to visualize the network structures \cite{Newman2004Finding}, enhance the information retrieval and promote the products recommendation \cite{Ding2016Prediction}, etc.. Social networks are typical attributed networks whose nodes are associated with attributes. Different from communities detected only based on structure cohesiveness property \cite{Fortunato2010Community,Xu2016Finding,Wu2015Multi}, communities with attribute subspaces \cite{Silva2012Mining,Gunnemann2013Spectral,Gunnemann2014Gamer,Huang2015Dense} not only have structure cohesiveness property, but also have attribute similarity property. An attribute subspace is a subset of the whole attribute dimension set. The subspace of a community should contain all attributes dimensions in which nodes in this community are similar to each other while different from nodes outside. Different communities usually have different subspaces \cite{Silva2012Mining,Gunnemann2013Spectral}. In the real world, applications usually require the set of communities whose subspaces containing specific concerned attributes rather than all communities in the network. These communities is called the application-aware community organization, and a subspace containing concerned attributes is called a feature subspace in this paper. It is noteworthy that besides the concerned attributes, each feature subspace may contain some other relevant attributes in which nodes in its embedded community are also similar to each other while different from nodes outside.

In this paper, we study the problem of mining the application-aware community organization w.r.t. some concerned attributes. Application-aware community organization is called community organization for short. The concerned attributes are designated in advance by a user based on the requirements of applications. Compared to the whole set of communities detected by traditional unsupervised clustering techniques \cite{Xu2016Finding,Wu2015Multi}, the specific community organizations are more suitable for applications. Take marketing as an example, a product is more likely to be prevalent in set of customer communities where people interact frequently and have some demand attributes about such product. When a merchant wants to advertise a product, he usually only have the basic idea of some demand attributes that the potential customers may have, while detecting the community organization w.r.t. these demand attributes will be great helpful for advertising such product. Take information propagation maximization in social networks as another example, the information with specific keywords is more easily to propagate in communities where people interact frequently and are interested in those keywords. When maximizing the propagation of a piece of information, its keywords are first extracted, and then the community organization with feature subspaces containing these keywords is ideal for releasing the information. Recently, several semi-supervised attribute community detection methods which can adjust the detection results based on user interests were proposed. FocusCO \cite{Perozzi2014Focused} aims to extract communities whose nodes are similar to the exemplar nodes that a user provides in advance. DCM \cite{Pool2014Description} mines communities with descriptions from a set of provided candidate communities. They are different from the problem of mining the application-aware community organization in this paper. 

To solve the proposed problem, we put forward ACM, an Application-aware Community organization Mining method. A set of concerned attributes is designated in advance. Since feature subspaces may contain some other implicit relevant attributes besides the designated ones, the set of designated concerned attributes should be expanded to complete feature subspaces so that the subspaces and their embedded communities match with each other. The set of concerned attributes is set as the initial subspace. In order to locate the potential communities belonging to the community organization, edges between nodes having large similarity in initial subspace are sampled to construct a network backbone and the cohesive parts of the network backbone are detected and set as the initial communities. Then each community and its subspace are adjusted based on each other so that the nodes inside the community are as similar as possible while as different from the nodes outside as possible in its subspace. The adjustment goals of each community and its subspace are consistent. Thus, they are adjusted iteratively by optimizing a unified quality function called subspace fitness. Finally, the redundant communities and subspaces are eliminated. 

The rest of this paper is organized as follows. Some related works are discussed in section 2. Section 3 describes and models the proposed problem. Section 4 presents the greedy ACM algorithm in details. Experiments results are analyzed in section 5. Finally, section 6 concludes the paper.

\section{Related Work}
Most of attribute community detection methods take unsupervised clustering techniques on link structure and node attributes simultaneously. In the early stage of the development, they require the nodes in each community to be similar in attribute full space \cite{Zhou2009Graph,Zhou2010Clustering,Cheng2011Clustering,Cheng2012Clustering,Akoglu2012Pics,Xu2012A,Yang2013Community,Ruan2013Efficient,Wu2016Multi}. The SA-Cluster \cite{Zhou2009Graph,Cheng2011Clustering} and its extended version Inc-Cluster \cite{Zhou2010Clustering,Cheng2012Clustering} define a unified neighborhood random walk distance on an augmented graph by considering all available attributes as additional attribute vertices, and then take a K-Medoids method to cluster the network based on this unified distance. CESNA method \cite{Yang2013Community} and BAGC method \cite{Xu2012A,Xu2014Gbagc} statistically model the link structure and all available node attributes, and then obtain communities by inferring parameters of their statistical models. PICS method \cite{Akoglu2012Pics} defines an encoding cost used to describe the adjacency matrix and attribute matrix of a network and gets communities by minimizing the cost. CODICIL method \cite{Ruan2013Efficient} creates content edges based on content similarity and combines content edges with structure edges. Then it samples edges that are locally relevant for each node and cluster the resulting backbone network by using standard community detection methods. With the increasing dimensionality of attribute space, the discrimination power of the attribute distance or similarity in full space may decrease \cite{Gunnemann2014Gamer}. Thus attribute subspace community detection methods \cite{Moser2009Mining,Silva2012Mining,Gunnemann2013Spectral,Gunnemann2014Gamer,Galbrun2014Overlapping,Huang2015Dense,Revelle2015Finding,Liu2015Community,Atzmueller2016Description,Yin2016Discovering} mining communities with nodes similar in attribute subspaces are preferred in most cases. CoPaM method \cite{Moser2009Mining} mines dense and connected subgraphs with homogeneous values in attribute subspaces by efficient pruning strategies. SCPM method \cite{Silva2012Mining} combines search and pruning strategies to mine structural correlation patterns which are dense subgraphs induced by attribute subspaces. SSCG method \cite{Gunnemann2013Spectral} adopts spectral clustering on an affinity matrix. It simultaneously learns communities and updates the affinity matrix by optimizing an objective function. GAMer method \cite{Gunnemann2014Gamer} defines twofold clusters by combing the paradigms of dense subgraph mining and attribute subspace clustering and mines them by various pruning strategies. SCMAG method \cite{Huang2015Dense} identifies cell-based subspace clusters composed of cells with dense coverage and connectivity in the subspace. Both of attribute full space and subspace methods mentioned above are unsupervised. Their results can not be guided based on the requirements of some application.

Two approaches using semi-supervised techniques have recently been proposed. FocusCO \cite{Perozzi2014Focused} allows user to steer the communities by providing a small set of exemplar nodes that are similar to one another as well as similar to the type of nodes the communities of his interest should contain. The application scenarios of their problem are different from ours, as we require a user to designate a set of concerned attributes which are contained by the mined subspaces. DCM \cite{Pool2014Description} aims to mine communities with descriptions from a set of candidate communities. The candidate communities can be inferred from designated descriptions. However, the community descriptions in DCM are different from general attribute subspaces in our method, as the descriptions are defined as queries composed of disjunctions of conjunctions over basic conditions. Moreover, one designated description can only obtain one initial candidate community, and in turn one final community in DCM, while one set of designated attributes can obtain an community organization in our method.  

\section{Problem Formulation}
In this section, the unified quality function of the community and its subspace is first defined and then the formal problem of mining application-aware community organization is given. An attributed network is defined as a 3-tuple $\mathcal{G}=(\mathcal{V},\mathcal{E},\mathcal{F})$, where $\mathcal{V}$ is a set of $n$ nodes, $\mathcal{E}$ is a set of $m$ edges and $\mathcal{F}: \mathcal{V}\rightarrow\mathcal{D}_1\times\cdots\times\mathcal{D}_r$ is an attribute function ($\mathcal{F}(v),v\in \mathcal{V}$ is an attribute vector of node $v$). $Dim=\{1,2,\cdots,r\}$ is the set of dimensions of attribute full space and $\mathcal{D}_r$ is the domain of values of attribute dimension $r$.

\subsection{Quality Function Definition}
In general, a community is a node subset densely intra-connected while sparsely connected to the rest of the network. Among multiple quality functions evaluating the structure cohesiveness of a single local community, we focus on the fitness \cite{Lancichinetti2009Detecting} because of its simple expression and good results. Let $A=[A_{u,v}]_{u,v=1}^n$ be the adjacency matrix of a network. The fitness of a community $C$ is defined as
\begin{equation}\label{E1}
  fit_A^C = \frac{invol_A^C}{vol_A^C},
\end{equation}
where $invol_A^C=\sum_{u,v\in C}A_{u,v}$ is the total internal degrees of community $C$, and $vol_A^C=\sum_{u\in C,v\in \mathcal{V}}A_{u,v}$ is the total degrees of community $C$. The fitness of a community gets larger when it has more edges inside while less edges across the boundary. Thus the maximization of the fitness can be used to guide the adjustment of a community in the plain network without attributes.

Besides structure requirements, a community with attribute subspace should also satisfy the attribute requirements that nodes in a community should be similar with each other while be different from the nodes outside. Attribute and structure requirements can be synthesized by regarding the attribute similarities of connected nodes as the edge weights. Then maximizing the total edge weights inside a community while minimizing the total edge weights across the boundary will satify structure requirements and attribute requirements, simultaneously. We adopt radial basis function kernels (RBF kernels) of attribute vectors as the attribute similarity, i.e., $s(\mathcal{F}(v),\mathcal{F}(u))=k(||\mathcal{F}(v)-\mathcal{F}(u)||)$, where $k(\cdot)$ is a RBF kernel and $||\mathcal{F}(v)-\mathcal{F}(u)||$ is the norm of $\mathcal{F}(v)-\mathcal{F}(u)$. In order to make the similarity increase as the norm decreases, the kernels are restricted to those having a non-positive derivative, such as the Gaussian, Epanechnikov or Exponential kernel \cite{Genton2002Classes}. As suggested in \cite{Gunnemann2013Spectral}, we adopt Exponential kernel $k_\theta (x)=e^{-\frac{x}{\theta}}$ where $\theta$ is a scaling parameter and $x$ is the norm, because it helps a unbiased subspace fitness computation discussed later. Since we aim to extract communities embedded in specific attribute subspaces, we use the subspace weighted Euclidean norm:
\begin{equation}\label{E2}
\begin{split}
    & ||\mathcal{F}(v)-\mathcal{F}(u)||_{l^D} \\
    =& \sqrt{(\mathcal{F}(v)-\mathcal{F}(u))^T diag(l^D)(\mathcal{F}(v)-\mathcal{F}(u))},
\end{split}
\end{equation}
where $l^D$ is a subspace vector corresponding to an attribute subspace $D$, and $diag(l^D)$ denotes a diagonal matrix whose main diagonal is $l^D$. The subspace vector satisfies the normalized condition, i.e., $\sum_{i=1}^r l_i^D =1,l_i^D\geq0$. The element $l_i^D$ represents the importance of attribute dimension $i$ in the subspace $D$. We simplify the problem by only considering the existence of attribute dimensions in the subspace while ignoring their relative importance. Thus for a subspace $D=\{i_1,\cdots,i_t\}$ (we use the set of attribute dimensions $D=\{i_1,\cdots,i_t\}$ to represent the subspace $D=\{\mathcal{D}_{i_1},\cdots,\mathcal{D}_{i_t}\})$, the importance weights of attribute dimensions inside are set equal, while outside are set as 0, i.e.,
\begin{equation*}
  l_i^D=\left\{
  \begin{array}{rcl}
  \frac{1}{t}, &&i\in D=\{i_1,\cdots,i_t\} \\
  0, &&otherwise
  \end{array}
  ,
  \right.
\end{equation*}
The attribute similarity is computed by applying Exponential kernel on subspace weighted Euclidean norm. Based on the attribute similarity, the network is projected to a subspace $D$ and re-weighted as $A^D=[A_{u,v}^D]_{u,v=1}^n$, where
\begin{equation*}
  A_{u,v}^D=k_\theta(||\mathcal{F}(u)-\mathcal{F}(v)||_{l^D})\cdot \mathbb{I}((u,v)\in \mathcal{E}),
\end{equation*}
where $\mathbb{I}$ is the indicator function indicating whether the expression inside is true or not. If we want to mine a community $C$ embedded in a subspace $D$, the fitness should be modified as the subspace fitness $fit_{A^D}^C$ by substituting original adjacency matrix with the re-weighted one. In this case, the maximization of the subspace fitness $fit_{A^D}^C$ not only drives the community to have more edges inside while fewer edges across the boundary, but also makes nodes inside the community more similar with each other while less similar with nodes outside under the subspace.

In the above, the subspace for a community is assumed to be known in advance. However, in our problem the subspace is unknown except that it must contain some concerned attributes. Given a community, its subspace should match with it, i.e., its subspace should make the nodes inside as similar as possible while as different from the nodes outside as possible. By setting the attribute similarities as the edge weights in network, the adjustment goal of subspace is similar to that of community. Thus the optimization of subspace fitness can also be used to adjust the subspace. In order to make the subspace fitness fairly evaluate the subspaces with different number of dimensions, the computation of subspace fitness should be unbiased w.r.t. number of dimensions, i.e., the subspaces with different number of dimensions should have comparable subspace fitness values. The weighted Euclidean norm is the bridge between attribute subspace and the subspace fitness, thus the subspace fitness is unbiased iff the weighted Euclidean norm is unbiased w.r.t. the number of dimensions. As suggested in \cite{Gunnemann2013Spectral}, the $z$-score normalized version of Euclidean norm is unbiased w.r.t. number of dimensions. Thus, we should substitute the Euclidean norm in subspace fitness with its $z$-score normalized version defined as
\begin{equation*}
\begin{split}
    & ||\mathcal{F}(v)-\mathcal{F}(u)||_{l^D}^z \\
    =& \frac{||\mathcal{F}(v)-\mathcal{F}(u)||_{l^D}-E[||\mathcal{F}(v)-\mathcal{F}(u)||_{l^D}]}{\sqrt{Var(||\mathcal{F}(v)-\mathcal{F}(u)||_{l^D})}}+Q,
\end{split}
\end{equation*}
where $Q$ is a constant to guarantee the $z$-score to be non-negative. With the similar proof procedure in \cite{Gunnemann2013Spectral}, it can be proved that when computing the subspace fitness, the Exponential kernel with scaling parameter $\theta$ using $z$-score normalized version of the weighted Euclidean norm is equivalent to the Exponential kernel with scaling parameter $\theta\cdot\sqrt{(Var(||\mathcal{F}(v)-\mathcal{F}(u)||_{l^D})}$ using original weighted Euclidean norm. Thus the $z$-score normalized version of the weighted Euclidean norm does not need to be computed in practice. Unbiased subspace fitness is guaranteed by using original Euclidean norm with the scaling parameter $\theta\cdot\sqrt{(Var(||\mathcal{F}(v)-\mathcal{F}(u)||_{l^D})}$. Since $\theta$ has little to do with our major work, it is set default as the recommended value, i.e., $\theta=1$.

\subsection{Problem Definition}
Having defined the subspace fitness, the problem of mining community organization is defined as follows.
\newtheorem{definition}{Definition}
\begin{definition}[Application-aware community organization mining problem]
  Given an attributed network $\mathcal{G}=(\mathcal{V},\mathcal{E},\mathcal{F})$, and the set of concerned attributes $D'=\{i_1,\cdots,i_t\}$, the application-aware community organization mining problem is to mine the set of community-subspace pairs $\mathcal{P}=\{(C,D)\}$, where $C\subseteq\mathcal{V}$, $D'\subseteq D\subseteq Dim$, and $(C,D)$ locally maximizes the subspace fitness:
  \begin{equation}\label{E3}
    fit_{A^D}^C=\frac{\sum_{u,v\in C}A_{u,v}^D}{\sum_{u\in C,v\in \mathcal{V}}A_{u,v}^D},
  \end{equation}
  where
  \begin{equation}\label{E4}
    A_{u,v}^D=e^{-\frac{||\mathcal{F}(u)-\mathcal{F}(v)||_{l^D}}{\theta\cdot\sqrt{Var(||\mathcal{F}(u)-\mathcal{F}(v)||_{l^D})}}}\cdot\mathbb{I}((u,v)\in \mathcal{E}).
  \end{equation}
\end{definition}
Locally maximizing the subspace fitness means that given a subspace $D$, neither adding any node to nor removing any node from $C$ can increase the subspace fitness, while given a community $C$, neither adding any attribute to nor removing any attribute from $D$ can increase the subspace fitness. 

Since mining each community-subspace pair independently optimizes its own subspace fitness, some community-subspace pairs which are too similar to some others are redundant in the mined community organization. Given two redundancy parameters $\beta_C, \beta_D\in[0,1]$, a community-subspace pair $(C',D')$ is redundant w.r.t. $(C,D)$ ($(C',D')\preccurlyeq_{red} (C,D)$), iff $fit_{A^{D'}}^{C'}\leq fit_{A^D}^C  \wedge  \frac{|C'\cap C|}{|C'\cup C|}\geq \beta_C  \wedge  \frac{|D'\cap D|}{|D'\cup D|}\geq\beta_D$. The final diverse community organization $\mathcal{H}$ is output by eliminating all redundant pairs. $\beta_C$, $\beta_D$ control the overlapping degree between different community-subspace pairs in the diverse community organization. We recommend $\beta_C=\beta_D=0.5$ to allow moderate overlaps if there is no specific preferences.

At least three types of attributes exist in real-world networks, i.e., numerical, binary and categorical attributes. The value differences of three attribute types are defined uniformly to make them be treated fairly. The value of any numerical attribute is normalized to the range $\mathcal{F}_i(v)\in[0,1]$ first. Then the difference is defined as $\mathcal{F}_i(v)-\mathcal{F}_i(u)$. The difference of a categorical attribute is set as 0 if two values are the same, otherwise 1. For a binary attribute, 1-0 indicate whether a node has the attribute or not. Thus the difference is set as 0 if two nodes have the attribute, otherwise 1. By the uniform definition, the differences of all types of attributes range from 0 to 1.

\section{Algorithm}
In the following section, we introduce our heuristic approximation algorithm ACM for the community organization mining problem. Since our problem is mining the community organization w.r.t. the concerned attributes, we first locate the communities potentially belonging to the community organization by constructing the set of initial community seeds. The set of concerned attributes is set as the initial attribute subspace for all community seeds. Then the set of community seeds and their subspaces are iteratively improved by two local expansion optimization processes. The overall algorithm will be first introduced. Then the key individual procedures will be explained in details.

\subsection{The ACM Algorithm}
The overall ACM algorithm is given in Algorithm 1. It first locates the required communities by constructing the set of community seeds. The concerned attributes guide the construction of community seeds. Since each required community is cohesive and has subspace containing the concerned attributes, the community seed should be cohesive and have similar values on all concerned attributes. A network backbone composed of edges between nodes having similar values on all concerned attributes is first constructed. Two values of an attribute are considered to be similar if their difference is smaller than $\pi$ percent of the average difference between values of any connected nodes. $\pi$ is a size parameter controlling the size of the network backbone. Experiments show that $\pi=1$ is appropriate to extract edges with significant large similarity. Thus $\pi$ is set as 1 if there is no specific preference. Cohesive parts of the network backbone are detected by a community detection algorithm LPA \cite{Raghavan2007Near}. All cohesive parts with size larger than a threshold are set as the community seeds.
\renewcommand{\algorithmicrequire}{\textbf{Input:}}
\renewcommand{\algorithmicensure}{\textbf{Output:}}
\begin{algorithm}
  \caption{ACM}
  \begin{algorithmic}[1]
    \REQUIRE attributed network $\mathcal{G}=(\mathcal{V},\mathcal{E},\mathcal{F})$, redundancy parameters $\beta_C$, $\beta_D$, size parameter $\pi$, and the set of concerned attributes $D'=\{i_1,\cdots,i_t\}$.
    \ENSURE the diverse community organization $\mathcal{H}$.
    \STATE $\mathcal{H}\leftarrow\emptyset$; $visitedNodes\leftarrow\emptyset$;
    \STATE $\mathcal{C}\leftarrow \mathrm{CONSTRUCT\_SEED\_SET}(\mathcal{G},D',\pi)$;
    \FOR {each $C\in\mathcal{C}$}
    \IF {$C\nsubseteq visitedNodes$}
    \STATE $D\leftarrow D'$;   
    \STATE $A^{D}\leftarrow \mathrm{REWEIGH}(\mathcal{G},D)$;
    \REPEAT
    \STATE $C\!\leftarrow\!\mathrm{ADJUST\_COMMUNITY}(A^{D},C)$;
    \STATE $(D,A^{D})\leftarrow\mathrm{ADJUST\_SUBSPACE}(A^{D},C,D)$;
    \UNTIL {$(C,D)$ unchanged}
    \IF {$D'\nsubseteq D$}
    \STATE continue;
    \ENDIF
    \STATE $\mathcal{H}\leftarrow\mathcal{H}\cup\{(C,D)\}$;
    \STATE $visitedNodes\leftarrow visitedNodes\cup C$;
    \ENDIF
    \ENDFOR
    \STATE $\mathcal{H}\leftarrow\mathrm{SELECT\_DIVERSE\_PAIRS}(\mathcal{H},\beta_C,\beta_D)$;
    \RETURN $\mathcal{H}$;
  \end{algorithmic}
\end{algorithm}

A simple heuristic strategy is taken to avoid redundant communities and unnecessary running time. Nodes in any mined community are labeled as visited nodes. If all nodes of a community seed are visited, such seed is discarded as it has large potential to grow to a redundant community. For each remaining community seed, its initial subspace is set as the set of concerned attributes. The network is first re-weighted based on initial subspace by equation (4). The community and the subspace are iteratively adjusted by two similar greedy hill-climbing techniques to maximize the subspace fitness. The adjustment procedures are described in the next subsection. The iterations do not stop until the community-subspace pair no longer changes. The iteration convergence is guaranteed, because both adjustment processes improve the subspace fitness and the subspace fitness has a maximum value 1. Since our problem aims to mine the subspaces containing the set of concerned attributes, the mined subspaces which do not contain the set of concerned attributes are discarded.

After all community seeds have been processed, a post-processing step is performed to select the diverse community-subspace pairs. The diverse community organization is initially set as empty. All mined pairs are sorted in descending order of their subspace fitness and considered one by one. If the considered pair is not redundant to any one in the diverse community organization, it is added to the diverse community organization. The diverse community organization is returned as the final output.

\subsection{Adjustment Procedures}
It is observed from the equation (3) that community and subspace jointly determine the subspace fitness, so the subspace fitness can be maximized by iteratively adjusting them. Inspired by the algorithm in \cite{Lancichinetti2009Detecting}, we take two similar greedy hill-climbing techniques to adjust the community and subspace, respectively. The pseudo-codes of two adjustment procedures are given in Procedure 1 and 2, respectively.
\setcounter{algorithm}{0}
\floatname{algorithm}{Procedure}
\begin{algorithm}
  \caption{$\mathrm{ADJUST\_COMMUNITY}$}
  \begin{algorithmic}[1]
    \REQUIRE re-weighted network adjacency matrix $A^D$, initial community $C$.
    \ENSURE improved community $C$.
    \REPEAT
    \STATE $\Delta f_{best}\leftarrow0$;
    \STATE $Actions\leftarrow\{\mathrm{REMOVE}(v)| v\in C\}\cup\{\mathrm{ADD}(v)|v\in \mathcal{V}\setminus C \wedge \exists u\in C:(v,u)\in \mathcal{E}\}$;
    \FOR {each $a\in Actions$}
    \STATE $\Delta f\leftarrow \mathrm{GET\_\Delta\_FITNESS}(A^D,a,C)$;
    \IF {$\Delta f>\Delta f_{best}$}
    \STATE $\Delta f_{best}\leftarrow \Delta f$; $bestAction\leftarrow a$;
    \ENDIF
    \ENDFOR
    \IF {$\Delta f_{best}>0$}
    \STATE $C\leftarrow \mathrm{MODIFY}(C,bestAction)$;
    \ENDIF
    \UNTIL {$\Delta f_{best}=0$}
    \RETURN $C$;
  \end{algorithmic}
\end{algorithm}

$\mathrm{ADJUST\_COMMUNITY}$ updates the community while fixes the subspace. It iteratively either adds a neighbor to or removes a node from the current community if such operation increases the subspace fitness. For a current community, a set of possible actions including removing nodes from it and adding neighbors to it is determined. The subspace fitness change of each action is calculated. The action leading to the largest positive change is selected to modify the community. The iteration continues until no action leads to the increase of the subspace fitness. The convergence is guaranteed, because each modification increases the subspace fitness and its maximum value is 1. The most time consuming step of $\mathrm{ADJUST\_COMMUNITY}$ is computing the fitness change of an action. For adding/removing a node $v$, the fitness change is calculated as
\begin{equation*}
  \Delta f=\frac{invol_{A^D}^C\pm wd_v^{in}}{vol_{A^D}^C\pm wd_v} -\frac{invol_{A^D}^C}{vol_{A^D}^C} ;
\end{equation*}
where $wd_v^{in}$ is the internal weighted degree of $v$ w.r.t. current community $C$, and $wd_v$ is the total weighted degree of $v$. Thus by saving the total internal degree $invol_{A^D}^C$ and the total degree $vol_{A^D}^C$, the fitness change can be calculated with time complexity $O(d_v)$, where $d_v$ is the degree of $v$. Accordingly, the calculation of subspace fitness of a community $C$ requires $O(|C|\cdot \overline{d})$ according to the equation (3), where $\overline{d}$ is the average node degree, and $|C|$ is the size of $C$.
\begin{algorithm}
  \caption{$\mathrm{ADJUST\_SUBSPACE}$}
  \begin{algorithmic}[1]
    \REQUIRE re-weighted network adjacency matrix $A^D$, initial community $C$, initial subspace $D$.
    \ENSURE improved subspace $D$, updated re-weighted network adjacency matrix $A^D$.
    \REPEAT
    \STATE $\Delta f_{best}\leftarrow 0$;
    \STATE $Actions\leftarrow\{\mathrm{REMOVE}(i)|i\in D\}\cup\{\mathrm{ADD}(i)|i\in Dim\setminus D\}$;
    \FOR {each $a\in Actions$}
    \STATE $(\Delta f,A^D_{temp})\leftarrow \mathrm{GET\_\Delta\_FITNESS}(A^D,a,C,D)$;
    \IF {$\Delta f>\Delta f_{best}$}
    \STATE $\Delta f_{best}\leftarrow\Delta f$; $bestAction\leftarrow a$; $A^D_{best}\leftarrow A^D_{temp}$;
    \ENDIF
    \ENDFOR
    \IF {$\Delta f_{best}>0$}
    \STATE $D\leftarrow \mathrm{MODIFY}(D,bestAction)$; $A^D\leftarrow A^D_{best}$;
    \ENDIF
    \UNTIL {$\Delta f_{best}=0$}
    \RETURN $D$ and $A^D$;
  \end{algorithmic}
\end{algorithm}

$\mathrm{ADJUST\_SUBSPACE}$ updates the subspace while fixes the community. Most of its steps are similar to those of $\mathrm{ADJUST\_COMMUNITY}$, except that it iteratively either adds an attribute to or removes an attribute from the current subspace. The most time consuming step of $\mathrm{ADJUST\_SUBSPACE}$ is also computing the fitness change of an action. Though the $\mathrm{GET\_\Delta\_FITNESS}$ has the same name as that in $\mathrm{ADJUST\_COMMUNITY}$, they have different steps inside. In $\mathrm{ADJUST\_SUBSPACE}$, the action changes edge weights which, in turn, changes the subspace fitness. For adding/removing an attribute $i$, the update of Euclidean norm can be calculated as
\begin{equation*}
  ||\cdot||_{l^{D^*}}=\sqrt{\frac{|D|\cdot||\cdot||_{l^D}^2\pm(\mathcal{F}_i(v)-\mathcal{F}_i (u))^2}{|D^*|}};
\end{equation*}
where $||\cdot||$ is short for $||\mathcal{F}(v)-\mathcal{F}(u)||$, $D^*$ is the update of $D$ after the action, and $|D|$ is the size of $D$. By saving $||\cdot||_{l^D}$, $||\cdot||_{l^{D^*}}$ can be computed in constant time. Thus by saving the current Euclidean norms of all edges, the updates of Euclidean norms and edge weights of all edges can be calculated with time complexity $O(m)$. After calculating the updates of edge weights, the update of fitness can be calculated according to equation (3) with time $O(|C|\cdot\overline{d})$.

\subsection{Computational Complexity}
Given an attributed network with $n$ nodes, $m$ edges and $r$ attributes, the time complexity of ACM is analyzed as follows.

$\mathrm{CONSTRUCT\_SEED\_SET}$ requires $O(m)$ to construct the network backbone and $O(m)$ to extract the seeds from backbone by a community detection algorithm with linear time complexity. For each community seed containing at least one unvisited node, $\mathrm{REWEIGH}$ takes $O(m\cdot r)$, $\mathrm{ADJUST\_COMMUNITY}$ takes $O(|C|\cdot n\cdot\overline{d})$, as performing each action averagely needs $O(\overline{d})$, at most $n$ actions are performed to adjust one node of the community and around about $|C|$ nodes need to be adjusted. $\mathrm{ADJUST\_SUBSPACE}$ requires $O(|D|\cdot r\cdot(m+|C|\cdot \overline{d}))$, as performing each action takes $O(m+|C|\cdot\overline{d})$, at most $r$ actions are performed to adjust one attribute dimension of the subspace and around about $|D|$ dimensions need to be adjusted. Assuming that the number of mined community-subspace pairs is $c$, all community-subspace pairs are sorted with time $O(c\cdot\log(c))$. Assuming that the size of final diverse community organization is $h$, each candidate pair is checked with all pairs in diverse community organization with time at most $O(h\cdot(\overline{|C|}+\overline{|D|}))$, where $\overline{|C|}$ and $\overline{|D|}$ are the average size of a community and a subspace, respectively. Thus $\mathrm{SELECT\_DIVERSE\_PAIRS}$ takes $O(c\cdot\log(c)+c\cdot h\cdot(\overline{|C|}+\overline{|D|}))$. Assuming that the average number of iterations required to converge a community-subspace pair is $q$, then the total time complexity is $O(m+c\cdot(m\cdot r+q\cdot(\overline{|C|}\cdot n\cdot \overline{d}+\overline{|D|}\cdot r\cdot(m+\overline{|C|}\cdot \overline{d})))+c\cdot\log(c)+c\cdot h\cdot(\overline{|C|}+\overline{|D|}))$. $q$ and $\overline{d}$ usually do not increase as the network size increases and are far smaller than the attributed network size, i.e., $n$, $m$, $r$. Thus they can be regarded as constants in time complexity. Meanwhile, $\overline{|C|}$, $\log(c)$ and $h$ are usually far smaller than $m$. Based on the operational rules of the symbol $O$, the total time complexity is simplified to $O(c\cdot(\overline{|C|}\cdot n+\overline{|D|}\cdot r\cdot m))$.

\section{Experimental Results}
In this section, we thoroughly evaluate the effectiveness and efficiency of ACM on synthetic networks and show its application values on real-world networks.

\subsection{Evaluation on Synthetic Networks}
The Experiments Settings are described first. Then the comparison results are analyzed.

\subsubsection{Experiments Settings}
The synthetic attributed networks with ground truth communities are generated based on the LFR benchmarks \cite{Lancichinetti2009Benchmarks}. Their degree and community size distribution are governed by power laws with exponents $\tau_1$ and $\tau_2$, respectively. The benchmarks are controlled by several parameters, i.e., node number $n$, average node degree $d_{avg}$, maximum node degree $d_{max}$, minimum community size $c_{min}$, maximum community size $c_{max}$ and mixing parameter $\mu$. Each node shares a fraction $1-\mu$ of edges with nodes in its own community and a fraction $\mu$ of edges with the rest of the network. The larger the $\mu$ is, the fuzzier the benchmark is. We attach three types of attribute vectors to each node to generate three types of attributed benchmarks, i.e., numerical, binary and categorical, respectively. Each method will run on the most suitable type of benchmark. The generation of attribute vectors is controlled by three parameters, i.e., attribute number $r$, attribute subspace size $t$ for each community and similarity probability $p$. All nodes of a community have similar attribute values in its subspace with probability $p$. The larger the $p$ is, the more homogeneous the community is in its subspace. The default parameter settings are as follows, $\tau_1=2$, $\tau_2=1$, $n=5000$, $d_{avg}=30$, $d_{max}=100$, $c_{min}=40$, $c_{max}=2c_{min}=80$, $\mu=0.2$, $r=20$, $t=6$, $p=0.9$. Six sets of benchmarks are generated by separately varying $n$, $c_{min}$, $\mu$, $r$, $t$, and $p$, while fixing all others.
\begin{figure}
\centering
\includegraphics{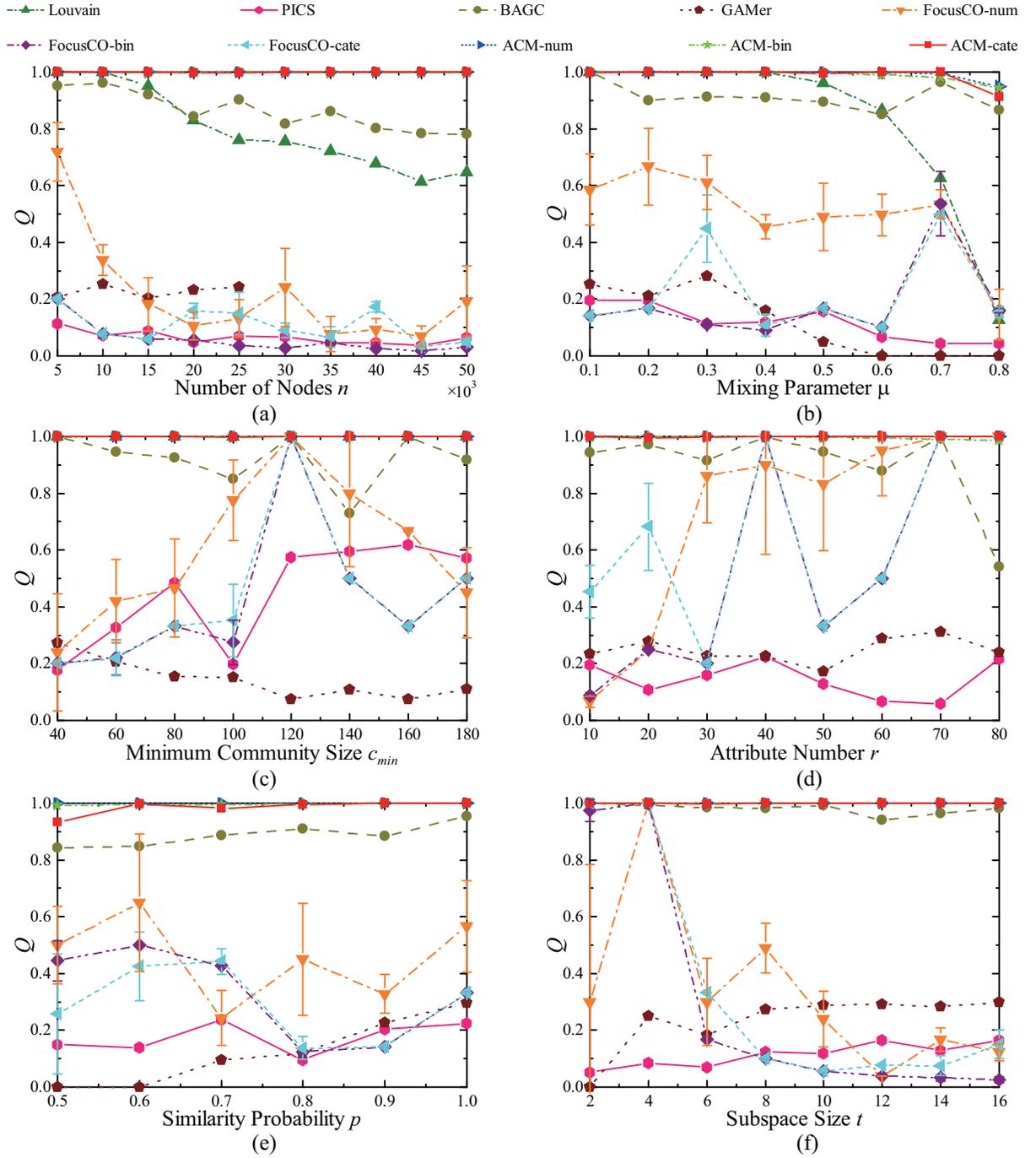}
\caption{$Q$ vs. (a) $n$, (b) $\mu$, (c) $c_{min}$, (d) $r$, (e) $p$, (f) $t$. Bars depict standard deviations. GAMer can't obtain results on networks larger than 25000, due to an out of memory problem.}
\label{F1}
\end{figure}

A variety of related methods are compared to ACM on synthetic networks. Louvain \cite{Vincent2008Fast} is an outstanding method only based on network structure. BAGC \cite{Xu2012A} and PICS \cite{Akoglu2012Pics} are two attribute full space methods. They run on categorical and binary benchmarks, respectively. BAGC requires the possible maximum community number as input. 1 to 5 times of real community number are set as input for BAGC respectively and the best result is reported. GAMer \cite{Gunnemann2014Gamer} is an attribute subspace method and it runs on numerical benchmarks. Finally, FocusCO \cite{Perozzi2014Focused} is a semi-supervised method. It controls the detected communities by providing a set of exemplar nodes. FocusCO is a randomized algorithm. Its average results and standard deviation over 20 runs are reported. FocusCO and ACM can run on all three types of benchmarks. Results on three types of benchmark are represented with a suffix '-num', '-bin', and '-cate', respectively. The parameters of other methods are set as default described in their papers.

In the experiments, we simulate the application scenarios of our problem. Since the goal of our problem is mining the community organization w.r.t. the concerned attributes, we randomly select two attributes as the concerned attributes in each benchmark. Once concerned attributes are determined, the ground truth community organization consists of ground truth communities whose subspaces containing these concerned attributes. In order to make the communities detected by FocusCO as similar to the communities in ground truth community organization as possible, 6 nodes with similar concerned attributes are selected as exemplar nodes for FocusCO. Other compared unsupervised methods can not use the concerned attributes to guide the detected communities. They partition the whole network. For every method, the detected communities which are the most similar to communities in the ground truth community organization are selected to evaluate its quality. The method quality is measured via an evaluation indicator $Q$ defined based on F1 score. Let $\mathcal{P}=\{P_i\}$ denote the set of communities in the ground truth community organization. Let $\mathcal{R}=\{R_j\}$ denote the set of communities detected by a method. The F1 score between each community $P_i$ in $\mathcal{P}$ and each community $R_j$ in $\mathcal{R}$, $F1(P_i,R_j)$, is computed. Since our goal is mining all communities in $\mathcal{P}$ accurately, the quality of each community $P_i$ in $\mathcal{P}$ is measured by the maximum F1 score between it and all detected communities in $\mathcal{R}$, i.e., $QI(P_i) = \max_{R_j\in \mathcal{R}}F1(P_i,R_j)$. The evaluation indicator $Q$ of a method is defined as the average quality of all communities in $\mathcal{P}$, i.e., $Q=\sum_{P_i\in \mathcal{P}}QI(P_i)/|\mathcal{P}|$. The larger the $Q$ is, the better the method solves the studied problem.

\subsubsection{Results Analysis}
The experiments are carried on six sets of benchmarks. Fig. 1(a) shows the results on benchmarks w.r.t. network size. Most of methods decrease their quality as $n$ increases. This is because when network gets larger, the size of ground truth community organization may increase and the problem becomes more challenging. However, ACM can mine the community organization perfectly in all cases. Fig. 1(b) shows the methods' quality change with mixing parameter. The quality of Louvain decreases a lot while that of others decreases not so much as $\mu$ gets larger. This illustrates that attribute information is beneficial to community detection in fuzzy networks. ACM nearly maintains the perfect quality even though $\mu$ reaches 0.8. Fig. 1(c) shows the methods' quality versus community size. The quality of GAMer decreases, as it tends to mine smaller communities. PICS has an increase tendency, as it tends to detect larger communities. ACM has the perfect quality all the time. Fig. 1(d) shows methods' quality for an increasing attribute number. ACM always has perfect quality. In Fig. 1(e), attribute similarity probability varies. ACM-num, BAGC and GAMer increase their quality as $p$ increases. ACM has the best quality in all cases. Finally, Fig. 1(f) shows the methods' quality versus subspace size. PICS increases its quality, as it is an full space method and larger subspace is closer to full space. ACM still has the perfect quality. In all six experiments, the performance of FocusCO has big oscillations. This is because the provided exemplar nodes in FocusCO cannot always infer all possible subspaces containing the set of concerned attributes. Thus it easily misses some communities in the ground truth community organization in many cases and its quality is oscillated. ACM has almost perfect quality all the time. This is because ACM can use the set of concerned attributes to accurately locate nearly all communities in the ground truth community organization, and it mines the subspace associated with each community which helps extract such community more accurately. The major reason for poor performance of other methods is that they are not specially designed for community organization mining problem, and can not adopt concerned attributes to guide their detection procedures.
\begin{figure}
\centering
\includegraphics{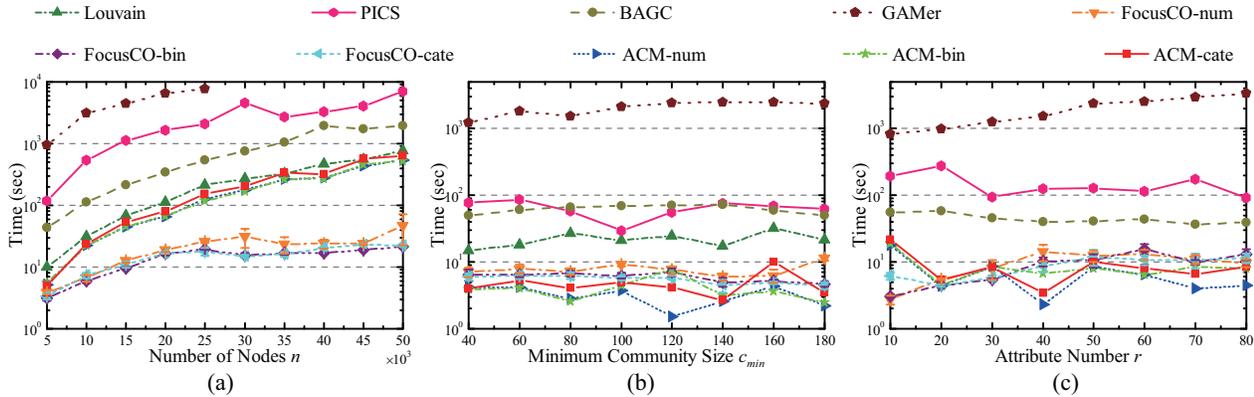}
\caption{Time vs. (a) $n$, (b) $c_{min}$, (c) $r$. Bars depict standard deviations.}
\label{F2}
\end{figure}

The methods' efficiency is shown in Fig. 2. In Fig. 2(a), all methods' running time increases as $n$ gets larger. In Fig. 2(b), the increase of community size makes the running time of GAMer increase, while it has little influence on other methods. Fig 2(c) shows the running time of GAMer increases a lot while that of others changes little as $t$ gets larger. In general, ACM and FocusCO are much faster than other methods, as they do not partition the whole network, but mine the required communities.

\subsection{Applications on Real-world Networks}
Since there is no ground truth about the community organization w.r.t. the concerned attributes in real-world networks, and other compared methods are not designed for community organization mining problem, it is inherently hard to quantitatively analyze our method on real-world networks. Our following case studies on real-world networks mainly illustrate the application values of ACM. IMDb \footnote{http://www.imdb.com/} \cite{Gunnemann2013Spectral} is a movie network where nodes represent movies with at least 200 rankings and an average ranking of at least 6.5. Two movies are connected if they share actors or if there exists a reference to each other. 21 movie genres are selected as attributes. Harvard \cite{Traud2012Social} is a Facebook network consisting of complete set of users from Harvard on one particular day in September 2005. Edges represent friendships among users. Each user is assigned with seven attributes, i.e., \emph{status flag}, \emph{gender}, \emph{major}, \emph{second major}, \emph{dorm}, \emph{year}, and \emph{high school}. arXiv \footnote{http://www.cs.cornell.edu/projects/kddcup/datasets.html} is a citation network where papers are represented by nodes and citations by edges. 400 keywords extracted from abstracts of papers are set as binary attributes and each attribute of a node indicates whether the specific keyword appears in the node's abstract or not. DBLP \footnote{http://dblp.uni-trier.de/} is a co-authorship network where nodes represent computer science authors and edges represent co-authorships. 21 conferences from five areas, database (DB), data mining (DM), information retrieval (IR), artificial intelligence (AI) and multimedia (MM), are set as attributes and the value of an attribute indicates how often the author has published papers in the specific conference. Authors attending at least one of the 21 conferences are included in the dataset. Dataset statistics and experiment statistical results are given in Table 1 and Table 2, respectively.
\begin{table}
\caption{Dataset Statistics. $n$: node number, $m$: edge number, $r$: attribute number.}
\label{T1}
\centering
\begin{tabular}{|c|c|c|c|}
\hline
Dataset & $n$ & $m$ & $r$ \\
\hline
IMDb & 862 & 4388 &	21\\
\hline
Harvard & 15126 &	824617 &	7\\
\hline
arXiv & 29555 &	352807 &	400\\
\hline
DBLP & 74373 &	586515 &	21\\
\hline
\end{tabular}
\end{table}

\begin{table}
\caption{Experiment Statistical Results. \#: number of community in mined organization, $\overline{|C|}$: average community size, Time: Running time, $\overline{f}$: subspace fitness averaged over mined communities. DBLP1 and DBLP2 are two results in DBLP with two different designated attributes sets, respectively.}
\label{T2}
\centering
\begin{tabular}{|c|c|c|c|c|}
\hline
Dataset & \# & $\overline{|C|}$ & Time (sec) & $\overline{f}$ \\
\hline
IMDb & 3 & 8.3 & 5.622e-1 &	1.000\\
\hline
Harvard &	234 & 15.3 &	3.632e+2 &	0.703\\
\hline
arXiv &	38 & 49.0 &	4.545e+3 &	1.000\\
\hline
DBLP1 &	23 & 57.2 &	2.750e+2 &	0.832\\
\hline
DBLP2 &	22 & 23.5 &	8.967e+1 &	0.890\\
\hline
\end{tabular}
\end{table}

In IMDb, the community organization w.r.t. attribute genres are mined. Assuming that a user wants to watch movies with genres Action and Adventure, \emph{action} and \emph{adventure} are set as concerned attributes. Three communities are mined in less than one second. The average community size is 8.3. The subspaces of two of them have genres \emph{action}, \emph{adventure} and \emph{sci-fi}, and that of the remaining one has genres \emph{action}, \emph{adventure} and \emph{family}. Due to the sparsity of networks and attributes, the subspace fitness values of all three communities are closed to 1. Thus the mined communities are densely intra-connected and sparsely connected with the rest of the network in the subspace re-weighted network. This case study shows ACM can help to recommend movies according to user's interest, and give the recommended movies a finer classification.

In Harvard, \emph{major} and \emph{dorm} are set as concerned attributes as we are assumed to mine communities having the same major and in the same dorm. 234 communities are extracted in about half minute. The average community size is 15.3. Besides some communities with subspace \{\emph{major},\emph{dorm}\}, other communities with subspaces \{\emph{major},\emph{dorm},\emph{second major}\}, \{\emph{major},\emph{dorm},\emph{year}\} and \{\emph{major},\emph{dorm},\emph{high school}\} etc. are also mined by ACM. Their subspaces include other relevant attributes to better describe them.

In arXiv, community organization w.r.t. some feature keywords are extracted. Assuming that we are interested in papers about theme Quantum Gravity, keywords \emph{quantum} and \emph{gravity} are set as concerned attributes. 38 communities about different aspects of Quantum Gravity have been mined in about 75 minutes. For example, a community of 35 papers concerning Quantization of Superstring has attributes \emph{quantum}, \emph{gravity}, \emph{superstring}, \emph{quantization}, \emph{fermion}, \emph{curve}, etc. A community of 25 papers concerning Lorentzian and Euclidean Quantum Gravity has attributes \emph{quantum}, \emph{gravity}, \emph{Lorentz}, \emph{space-time}, \emph{Euclidean}, \emph{integral}, etc. A community of 79 papers concerning Loop Quantum Gravity has attributes \emph{quantum}, \emph{gravity}, \emph{loop}, \emph{nonperturbative}, \emph{perturbative}, \emph{flat}, etc. Due to the sparsity of networks and attributes, the subspace fitness values of all mined communities are closed to 1 which means the mined communities are cohesive and well separated from the rest of the network in the subspace re-weighted network. Thus ACM can be used to analyze citation network and recommend relative papers with keywords pertaining to user interest. Moreover, it can group papers about different aspects of designated theme into different communities and give each of them a keyword subspace which describes the theme in more details.

In DBLP, assuming that we want to advertise some journals about data mining to researchers, conferences \emph{ICDM} and \emph{SIGKDD} from data mining are set as concerned attributes. DBLP1 in TABLE 2 records the experiment statistical results in this situation. 23 communities with average community size 57.2 are mined in about 5 minutes. All extracted subspaces contain some conferences from data mining field. Some of them also contain conferences from other fields. For example, one subspace has attributes \emph{ICDM}, \emph{SIGKDD}, \emph{PKDD}, \emph{SIGIR}. The first three conferences are from data mining and the last one is about information retrieval. Another subspace has attributes \emph{ICDM}, \emph{SIGKDD}, \emph{PKDD}, \emph{SDM}, \emph{NIPS}. The first four conferences are from data mining and the last one is from artificial intelligence. A subspace containing conferences from different areas is meaningful, as different areas in computer science overlap heavily. The journals about data mining can be advertised in the mined 23 communities where authors are likely to be interested in data mining. We also mine another community organization w.r.t. concerned attributes \emph{MM} and \emph{SIGGRAPH} from multimedia. DBLP2 in TABLE 2 records the experiment statistical results in this situation. 22 communities with average community size 23.5 are mined in about 1.5 minutes. The community organization with concerned attributes \emph{MM} and \emph{SIGGRAPH} is different from that with concerned attributes \emph{ICDM} and \emph{SIGKDD}. The average community size in former organization is smaller than that in the latter one. This may be because the academic circles in multimedia are smaller than those in data mining.

\section{Conclusion}
In this paper, we study an application oriented problem of mining application-aware community organization w.r.t. concerned attributes in social networks. A unified quality function called subspace fitness is defined to evaluate the quality of both subspace and community. A greedy algorithm ACM is developed to mine the application-aware community organization by optimizing the quality function. The experiments demonstrate the effectiveness and efficiency of ACM and show its application values.

\section*{Acknowledgment}
This work was supported by the National Natural Science Foundation of China (U1636105), the National Key Basic Research Program of China (2013CB329603).

\section*{References}

\bibliography{Manuscript}

\end{document}